# Efficacy of Put and Spd sprayed on leaves from *Brassica juncea* plants against $Cd^{2+}$-induced oxidative stress.

Michel AOUN[a], Annick HOURMANT[a] , Jean-Yves CABON[b]

Laboratoire de Toxicologie alimentaire et cellulaire, Université de Bretagne Occidentale (Brest-France)[a].

Laboratoires de Chimie, éléctrochimie moléculaire et chimie analytique-UMR 6521, Université de Bretagne Occidentale (Brest-France)[b].

*Dr. Michel AOUN, UBO
6, Avenue Le Gorgeu
29200 BREST
michel.aoun@univ-brest.fr*

**Abstract**

The protective effect exerted by polyamines (Put and Spd) against cadmium (Cd) stress was investigated in *Brassica juncea* plants. Treatment with $CdCl_2$ (75 µM) resulted in a rise of Cd accumulation, a decrease of fresh and dry weights in every plant organ, an increase of free polyamine content at limb and stem levels as well as a decrease at root level. On the other hand, the total conjugated polyamine levels in the stem tissues were unaffected by Cd. In the leaf tissues, this metal caused a reduction of chlorophyll a content, a rise of guaiacol peroxidase (GPOX) activity and an increase of malondialdehyde (MDA), soluble glucide, proline and amino acid contents.

Exogenous application, by spraying, of putrescine (Put) and spermidine (Spd) to leaf tissues reduced $CdCl_2$-induced stress. These polyamines proved to exert a partial, though significant, protection of the foliar fresh weight and to alleviate the oxidative stress generated by Cd through reductions of MDA amounts and GPOX (E.C.1.11.1.7) activity.

The enhancement of chlorophyll a content in plants by Put and those of Chl a and Chl b by Spd both constitute evidences of their efficacy against the $Cd^{2+}$-induced loss of pigments. Conversely to Put, Spd caused a decrease of Cd content in leave tissues and a rise in the stems and roots; these findings are in favour of a stimulation of Cd uptake by Spd. The proline stimulation observed with Cd was reduced further to the spraying of Put onto tissues, but the decrease induced by Spd was more limited. In the plants treated with Cd, the amino acid contents in the leaves were unaffected by Put and Spd spraying; on the other hand, $Cd^{2+}$ disturbe polyamine levels (free and acido-soluble conjugated-forms); we notice the rise of total free PAs and the decrease of their conjugated-ones.

The results of the present study suggest that Put and Spd can efficiently protect *Brassica juncea* plants against Cd-induced oxidative damage likely through a strong reduction of $H_2O_2$ generation and pigment protection. Moreover, the use of Spd in phytoremediation processes is briefly discussed.

**Abbreviations**

| | |
|---|---|
| GPOX | Guaiacol peroxidase |
| MDA | Malondialdehyde |
| MES | 2-(N-morpholino ethane sulfonic) acid |
| PAs | Polyamines |
| PCA | Perchloric acid |
| Put | Putrescine |
| Spd | Spermidine |
| Spm | Spermine |
| SSAT | Spermidine / Spermine N1-Acetyltransferase |

**Introduction**

The contamination of soils, water and the atmosphere further to the dispersal of industrial and urban wastes generated by human activities is, nowadays, a major environmental concern. It is caused by a wide range of organic contaminants (combustible and putriscible substances, hazardous wastes, explosives and petroleum products) and inorganic pollutants (heavy metals) (Adriano, 1986; Alloway, 1990).

Among these metals, Cd as well as Zn, Cu, Pb, …. are widespread contaminants known to be toxic to humans, animals and plants. Furthermore, $Cd^{2+}$ is easily taken up by plants, where its toxicity becomes manifest through growth reduction (Chen and Kao, 1995), decrease in chlorophyll contents (Stobart et al. 1985; Larsson et al. 1998) and decline in transpiration rate (Haag-Kerwer et al. 1999). Moreover, lipid peroxidation (Ferrat *et al.,* 2003) and guaiacol peroxidase activity (Shah *et al.,* 2001; Srivastava *et al.*, 2004; Smeets *et al.,* 2005; Mishra *et al.*, 2006; Hsu et Kao, 2007; Semane *et al.,* 2007) are enhanced by this metal and others, whereas carbohydrate metabolism (Moya *et al.,*1993; Costa and Spitz, 1997), amino acid content (Sharma and Dietz, 2006), proline (Lesko and Simon-Sarkadi, 2002; Singh and Tewari, 2003; Sharma and Dietz, 2006) and polyamine one (Lesko and Simon-Sarkadi, 2002; Groppa et al. 2003; Kuthanova et al. 2004; Taulavuori *et al.*, 2005 ; Sharma and Dietz, 2006; Groppa *et al.*, 2007a, b; Groppa and Benavides, 2008) are altered.

Oxygen is essential for aerobic organisms, but the reactive oxygen species (ROS), including the superoxide anion ($O_2^{\cdot -}$), hydroxyl radicals ($OH^{\cdot}$) and hydrogen peroxide ($H_2O_2$) (Foyer et al. 1994, 1997; Asada, 1999), which are generated in the course of metabolic reactions within the cells, are known to be toxic. Since long, ROS have been considered as damaging to cell membranes (Apel and Hirt, 2004) and implicated in ageing and wounding (Thompson et al. 1987). It is only recently that they have emerged as ubiquitous signalling molecules involved in the recognition of stress factors as well as in the response to them (Foyer and Noctor, 2005). Plants cope with the oxidative stress by using antioxidant enzymes such as superoxide dismutase (SOD, E.C. 1.15.1.1), ascorbate peroxidase (APX, E.C. 1.11.11.1), catalase (CAT, EC 1.11.1.6), glutathione reductase (GR, E.C. 1.6.4.2), peroxidase (POX, E.C.1.11.1.6), guaiacol peroxidase (GPOX, E.C. 1.11.1.7), glutathione peroxidase (GSH-PX, E.C. 1.11.1.9) and many NADP-dehydrogenases as well as antioxidant compounds of low molecular weight, such as ascorbate and glutathione (Noctor and Foyer, 1998; Asada, 1999; Del Rio et al. 1998, 2002).

The toxicity of Cd to plant cells is related to the oxidative stress caused through the generation of ROS (Piqueras et al. 1999; Sandalio et al. 2001; Schützendübel and Polle, 2002; Olmos et al. 2003; Kuo and Kao, 2004; Romero- Puertas et al. 2004), the stimulation, or inhibition, of the activities by antioxidant enzymes (Shaw, 1995; Gallego et al. 1996; Chaoui et al. 1997; Dixit et al. 2001; Shah et al. 2001; Innelli et al. 2002; Kuo and Kao, 2004), the production of oxidative damage, the induction of lipid peroxidation (Shaw, 1995; Gallego et al. 1996; Chaoui et al. 1997; Lozano-Rodriguez et al. 1997; Dixit et al. 2001; Shah et al. 2001, Chien et al. 2002; Kuo and Kao, 2004) and protein oxidation (Pena et al. 2006).

Aliphatic polyamines are synthesised in both Prokaryotes and Eukaryotes. Among them, putrescine (Put), spemidine (Spd) and spermine (Spm) are the most common ones. Moreover,

Put and Spd are usually more abundant than Spm, which is often present at trace amounts. In plants, these polyamines are not only involved in numerous cellular and molecular processes (Slocum et al. 1984; Bouchereau et al. 1999; Wallace et al. 2003), but also in their responses to environmental stresses by salt (Sun et al. 2002; Zhao et al. 2003), water (Xing et al. 2000), chilling (Lee, 1997; Zheng et al. 2000), acid rain (Velikova et al. 2000), ozone (Ormord and Beckerson, 1986; Navakoudis et al. 2003), paraquat (Ye et al. 1997) and heavy metal(s) (Groppa et al. 2001; Wang et al., 2004; Taulavuori et al. 2005; Sharma and Dietz, 2006; Groppa et al. 2007a, b). Moreover, in engineered plants, tolerance to multiple environmental stresses is enhanced by an overproduction of polyamines (Roy and Wu, 2001; Capell et al. 2004; Kasukabe et al. 2004). Less is known about the biosynthetic activities that lead to their accumulation in response to heavy metals (Weinstein et al. 1986; Wettlaufer et al. 1991; Groppa et al. 2003; Balestrasse et al. 2005).

Evidences of the protection exerted by exogenous polyamines against oxidative stress of various origins are available in the literature: they are about paraquat, a viologen often used as herbicide (Bors et al. 1989; Minton et al. 1990; Chang and Kao, 1997; Kurepa et al. 1998; Benavides et al. 2000), acid rain (Velikova et al. 2000), salt (Tang 123 and Newton, 2005) and heavy metals, e.g. Cd and Cu, (Groppa et al. 2001; Urano et al. 2003; Tang et al. 2005; Franchin et al. 2007; Groppa et al. 2007a, b; Hsu and Kao, 2007). Drolet and co-workers (1986) showed that polyamines are effective radical scavengers in a number of chemical- and *in vitro* enzyme-systems. Under osmotic stress, ROS production is enhanced by a reduction of polyamine content in the leaves of *Glycyrrhiza inflata* (Li and Wang, 2004).

These considerations led us to carry out experiments aimed at determining whether the toxicity by Cd was affected by spraying Put and Spd onto the leaves of *Brassica juncea* plants issued from seedlings grown hydroponically. They showed that polyamine metabolism was disturbed by Cd and that, further to an exogenous application of Put and Spd by spraying at the leaf tissues, the oxidative damage caused by $CdCl_2$ was reduced.

## Material and methods

Plant material

The cultivar, *Brassica juncea* I39/1, is a pure spring line, genetically fixed and obtained by autofertilisation.

Culture conditions

Indian mustard (*Brassica juncea*) seeds were grown on humidified vermiculite at 22°C for 3 days under dark conditions. At day 3, uniformly germinated seeds were selected and cultivated again on vermiculite for 15 days. Then, at day 18, plantlets were extracted from vermiculite and their roots were carefully washed under water flow. The extacted plants were grown, for 84 hours, in a hydroponic system composed of a half-strength Hoagland solution (Hoagland and Arnon, 1950) in MES buffer (1 mM) and enriched with $CdCl_2$ (75 µM). the final pH was 6.60. Solutions of Put (1 mM) and Spd (1 mM) were prepared in Teepol (0.01 %, v/v) and further sprayed at the surface of leaves. The control plants were sprayed with only Teepol. All experiments were carried out in a growth chamber maintained at 22°C (light)/20°C (dark) under a photoperiod (16 h) with 60 µmol photon.$m^{-2}.s^{-1}$) provided by cool white fluorescent lamps.

Determination of pigment contents

The contents in chlorophylls a, b and carotenoids were determined by using the procedure developed by Lichtenthaler (Lichtenthaler, 1987). Briefly, about 1 g of limb fresh weight was extracted at 4°C with acetone 100% (25 ml), then the mixture was centrifuged 10 min at 5 000 g and 4°C prior to the recording of absorbance at 470, 662 and 645 nm, respectively, with a UV-1605 spectrophotometer from Shimadzu.

Measurement of Cd in the different organs

Cadmium concentrations in the different solutions were measured by electrothermal atomic absorption spectrometry with a Perkin-Elmer, SIMAA 6100, spectrometer operated in the single-element mode. At harvest, the leaves, stems and roots were washed twice with distilled water for 5 min, weighed and then oven-dried for 4 days at 80°C prior to grounding. All solutions were prepared with high-purity water from a MilliQ-system (Millipore, Milford, MA, USA). Aliquots of about 200 mg of grounded matter were transferred to Teflon vessels prior to the addition of nitric acid, hydrogen peroxide and hydrofluoric acid in mixture (4/3/1, v/v/v). Then, the vessels were closed and exposed to microwave digestion as detailed elsewhere (Weiss et al. 1999).

Assessment of lipid peroxidation

Lipid peroxidation was assessed from the amount of produced malondialdehyde (MDA) able to react with thiobarbituric acid and was measured according to themethod developed by Minotti and Aust (1987) and improved by Iturbe-Ormaetxe and co-workers (1998).

Approximately 1 g of leaves (fresh weight) was extracted by grinding with 5 mL of a mixture of meta-phosphoric acid (5%, w/v) and butylhydroxy-toluene (50/1, v/v). After centrifugation at 5 000 g for 30 min, 4 ml of the supernatant were homogenized with 200 µL of butyhydroxytoluene (2%, w/v), 1 mL of HCl (25%, v/v) and 1 mL of 2-thiobarbituric acid (1%, w/v) prepared in NaOH (50 mM). The resulting mixture was incubated at 95°C for 30 min, then quickly cooled in an ice bath to stop the reaction. Finally, the chromogen was extracted by addition of 3 ml of 1-butanol to the mixture. Then, after centrifugation at 500 g for 5 min, the absorbance was read at 532 nm. The extinction coefficient of MDA ($\varepsilon$ = 155 mmol $L^{-1}$ $cm^{-1}$) was used to calculate the sample content in MDA.

Determination of guaiacol peroxidase activity (GPOX)

GPOX activity in leaves was measured by the procedure developed by MacAdam and co-workers (1992) after slight changes: briefly the use of guaiacol (4 µmol) as substrate by peroxidases in the presence of $H_2O_2$ leads to the formation of tetraguaiacol (1 µmol), whose absorption, read at 470 nm, gives the amount of enzyme. The extinction coefficient of tetraguaiacol used to calculate GPOX activity was 26.2 $mM^{-1}$ $cm^{-1}$. One unit of GPOX was defined as the amount of enzyme that caused the formation of 1 µmol tetraguaiacol per min. GPOX activity was expressed as the amount of oxidized guaiacol.

Amino acid determination

This determination is based on the fact that every amino acid reacts with ninhydrin at 100°C to form the Ruheman purple (Yemm and Cooking, 1955). Approximately, 1 g of leaves (fresh weight) was mixed in a glass tube with distilled water (7 mL) and heated at 100°C for 1 hour.

After cooling on ice, the mixture was ground and centrifuged at 5000 g for 10 min at 4°C. The pellet was submitted to a second extraction and centrifugation under the same conditions. Then, after pooling, 200 µl of supernatant were mixed with 0.5 ml of citrate buffer (160 mM, pH 4.6), before addition of 1 mL of ninhydrin (54 mM, w/v) dissolved in ethanol 70 % and 2 mL of ascorbic acid (1%, w/v). After strong agitation (vortex), the tubes were placed in a water bath at 100°C for 20 min, then cooled on ice prior to the addition of 3 mL of ethanol 70%. Then absorbance was read at 570 nm. The standard curve was constructed from a dilution series of a stock solution of L-leucine (5 mM). The amounts of amino acids were expressed as µmol equivalent leucine per g (fresh weight).

Measurement of free proline

The sample content in free proline was determined by spectrophotometry at 520 nm according to the procedure by Bates and co-workers. (1973): 1 g of leaves (fresh weight) was extracted with 10 mL of sulfosalycilique acid (3%, w/v). After filtration, 0.5 mL of the filtrate were incubated at 100°C with 1 mL ninhydrin (1%, w/v). The reaction was let to develop for 1 hour, and then stopped by cooling in an ice bath prior to the addition of 3 mL of toluene. After vortexing of the final mixture for 2 min, it was incubated for 3 h in the dark and at room temperature before reading of the absorbance at 520 nm. The standard curve was constructed from a dilution series of a stock solution of L-proline (5 mM). The amounts of proline were expressed in µmol.$g^{-1}$ (fresh weight).

Sugar determination

Soluble sugars were determined according to the method by Cerning-Beroard (1975): under certain conditions of pH and heat, soluble sugars are degraded into furfural, which further reacts with anthrone to form a blue compound whose absorbance is read at 625 nm. About 1 g (F.W) of leaf was mixed in a tube with 7 ml distilled water before heating at 100°C for one hour. After cooling on ice, the mixture was ground, then centrifuged for 10 min at 5 000 g and 4°C. The pellet was subjected to a second extraction and centrifugation under the same conditions. Then, the two supernatants were pooled, mixed and filtered on a 45-µm filter paper.

An aliquot (500 µl) was vortexed in glass tubes with 5 ml of a solution of anthrone (92 mg anthrone and 92 mg de thiourea dissolved in 100 ml sulfuric acid (70 %, v/v), then the mixture was heated at 100°C for 10 min. After cooling in an ice bath, absorbance was read at 625 nm. The results are expressed as µmoles equivalent glucose.$g^{-1}$ F.W.

Polyamine determination

Free PAs were extracted from leaf, stem and root tissues by the procedure described by Flores and Glaston (1982). To determine the amount of conjugated PAs in the acid soluble fraction, 200 µl of the supernatant were mixed with an equal volume of 12 N HCl and hydrolysed for a night at 110°C in flame-sealed ampoules. The hydrolysates were dried under a flow of air and resuspended in 200 µl PCA (5 %, v/v). About PAs, they were derivatised according to the method by Flores and Glaston (1982), slightly modified by Féray and co-workers (1992). The dansyl derivatives of putrescine (Put), spermidine (Spd) and spermine (Spm) were separated by HPLC and quantified by fluorescence spectrophotometry. The HPLC apparatus consisted of two solvent pumps (Kontron 422) coupled to a high performance mixer, a 7125 Rheodyne injection valve fitted with 20 µl loop and a stainless precolumn spheri ODS 5 mm (Brownlee (30 X 4.6 mm) in conjunction with a stainless steel analytical column (Ultrasphere). The mobile phase consisted of 0.01 M $NaH_2PO_4$, pH 4.4 as solvent (A) and of methanol/acetonitrile in mixture (50/50, v/v) as solvent B. Both solvents were mixed at ambient temperature to produce a gradient elution at a flow–rate of 1 $mL.min^{-1}$ starting with B 80% for 5 min, followed with B 80-89% for 2 min, then B 89-100% for 5 min, B 100-80% for 3 min and finally B 80% for 10 min. The gradient and data analysis were controlled by a microcomputer data system (Kontron Inst.). Detection by spectrofluorimetry was achieved with a SFM 25 apparatus (Kontron) set at 360 and 510 nm wavelengths for excitation and emission, respectively.

The retention times (min) for Put, Spd and Spm were 7.6, 13.5 and 15.4, respectively.

Statistical analysis

The different results are expressed as the mean ± SD (standard deviation). Three independent experiments were performed and each measurement was made in triplicate. The statistical differences between measurements were assessed by one-way ANOVA and LSD tests.

**Results**

By comparison with the controls, Figure 1 shows clearly the significant reductions in the fresh weight of organs from 18-day-old *Brassica juncea* plants (leaves: 34%, roots: 30% and stems: 20%) induced by the presence of $CdCl_2$ (75 µM) in the culture medium. One should note that the spraying of leaf tissues with Put or Spd in plants subjected to $CdCl_2$ had a positive effect by reducing the loss of fresh weight for the leaves (22-23% against 34%). On the other hand, this beneficial action was not observed for roots and stems (Fig. 1).

Table 1 highlight that 84 h-exposure of plants to $CdCl_2$ (75 µM) caused significant reductions in the dry weight (17% whatever the organ) and total water content of leaves (35%), roots (32%) and stems (24%). The dry weights of leaves and stems proved to be unaffected by spraying of Put or Spd upon the leaf tissues of the Cd-exposed plants conversely to that of the roots where the loss was only 9% against 17%.

Further to their exposure to $CdCl_2$, the plants accumulated substantial amounts in their organs. As expected, this accumulation was the greatest in the roots (4506 $\mu g.g^{-1}$ D.W.) and far much lower in the stems and leaves (933 and 355 $\mu g.g^{-1}$ D.W, respectively) (Table 2). Spraying of Put on leaf tissues caused no significant difference 303 in the Cd contents of organs, conversely to that of Spd: indeed, Cd contents was significantly decreased in the leaves (20%) and significantly increased in the stems 305 and roots (22 and 86 %, respectively).

Figure 2 shows that, by comparison to controls, exposure of plants to $CdCl_2$ (75 µM) caused a significant reduction of the leave contents in chlorophyll a (15%), but not in chlorophyll b and carotenoids ($P < 0.05$). Then the Chl a/b ratio in Cd-exposed plants was 2.84 against 3.04. Spraying of Put or Spd upon control plants had no significant effect upon the pigment contents (Fig. 2). It is worth noting that spraying of Put upon Cd-exposed plants resulted in an elevation of Chl a content. That of spd caused significant rises of Chl a and Chl b contents ($p < 0.05$), and, thus, the Chl a/b ratio was significantly lowered (2.34 against 2.84, control value: 3.04). However, the carotenoid contents was unaffected by spraying of Put or Spd.

Figure 3 evidences the enhancement of lipid peroxidation (+ 157%) by exposure to Cd. On the other hand, MDA contents were unaffected by spraying of Put or Spd onto control plants (Fig. 3). The MDA content of Cd-exposed plants was lowered further to the spraying with Put to control value but Spd proved to be less efficient (reduction of MDA content by 33% , though still 71% higher than in controls.

Exposure to $CdCl_2$ induced a significant enhancement of GPOX activity (2.4-fold with respect to controls, $p < 0.05$) (Fig. 4). On the other hand, it was unaffected by spraying of Put or Spd upon controls conversely to Cd-exposed plants where Put and Spd alleviated partially, but significantly the oxidative stress generated by Cd (1.9-fold reduction of GPOX activity).

Concerning the contents in free amino-acids, it was significantly increased (65%) by exposure to $CdCl_2$ (Fig. 5), but spraying of controls and Cd-exposed plants with Put or Spd had no significant effect upon the content in free amino-acids.

Table 3 evidences the rise in the proline content of leaf tissues (228%) induced by exposure to $CdCl_2$. This content was unaffected by spraying of control plants with Put or Spd, conversely to the case of Cd-exposed plants, where further to Put spraying the proline content was markedly lowered (113% against 228%). However, Spd proved to be less efficient (1.86 fold less than in Cd-exposed plants, but 80% higher than in control plants.

Figure 6 evidences the marked rise (117%) of soluble sugar levels observed further to the exposure of *Brassica juncea* seedlings to $CdCl_2$. Spraying of Put or Spd upon the leaves of control plants had no effect on their soluble sugar content. The high sugar levels observed in Cd-exposed plants were greatly lowered by Spd, but not by Put: indeed, they were 56% higher than in controls.

In comparison to control plants, exposure to $CdCl_2$ (75 μM) induced a significant accumulation of Put, Spd and Spm in the leaves (62, 33 and 31%) and the stems (13, 18 and 45%) (Fig. 7 A, B). On the other hand, in the roots, free polyamine levels were decreased by 22, 58 and 71% for Put, Spd and Spm, respectively (Fig. 7C). Put amounts ($P < 0.05$) were significantly increased at leaf and stem levels (Fig 7 A, B) and decreased in roots (Fig. 7 C) further to the spraying of Put (1 mM) on Cd-treated plants; moreover, Put (1 mM) decreased significantly Spd and Spm contents in leaf and root tissues (Fig. 7 A, C), but Spd and Spm stem levels were similar to those reported in Cd-treated plants (Fig. 7 B).

Spd (1 mM) applied exogenously increased significantly free Put contents at foliar, stem and root levels in comparison to Cd treated plants (Fig. 7 A, B, C). Moreover, Spd contents were increased significantly at foliar and root levels. However, Spd did not modified Spd stem levels and Spm contents at different organs in comparison to Cd treated plants (Fig. 7).

$CdCl_2$ (75 µM) treatment enhance significantly endogenous conjugated Put levels of 101 and 10 % at foliar and stem levels respectively (Fig. 8 A, B), but not Put root levels in comparison to control plants (Fig. 8 C). Cd did not affected conjugated Spd levels but increase conjugated Spm levels of 36 % at foliar levels (Fig. 8 A). However, Cd decrease Spd and Spm levels of 25 % at stem level, and 35 and 30 % at root levels in comparison to control values (Fig. 8 B, C). Put (1 mM) applied exogenously did not showed any significant effect on Put, Spd and Spm contents at foliar level ($P < 0.05$). However, Put and Spd contents were increased but not Spm at stem level in comparison to Cd treated plants. Moreover, Put contents were unchanged at roots level, but Spd and Spm contents were decreased in comparison to Cd treated plants. Spd (1 mM) applied exogenously did not show any significant effect on Spm contents but it was responsible for Put decrease and Spd increase at foliar level. At stem level, Spd application did not modify the contents in endigenous Spd and Spm, but cause a significant decrease of Put level ($P < 0.05$). At root level, Spd application didn’t have any effect on Spm contents, but decrease significantly Put and Spd contents at room levels in comparison to Cd-exposed plants.

**Discussion and conclusions**

The results showed that the application of $CdCl_2$ (75 µM) in hydroponic conditions, causing a decline in overall growth of plants du to reduced masses of fresh and dry materials of the various organs. The percentages of water are also down. These results agree with previous work on *Brassicacea* (Larsson et al. 1998; Haag-Kerwer et al. 1999; Singh and Tewari, 2003). The foliar application of the Put or the Spd does not affect the biomass of various organs of B. juncea but produced a slight increase in weight of dry matter accompanied by a slight decrease in the percentage of water from aerial parts.

When the foliar spraying with one or other of PAs was associated with exposur to $CdCl_2$ (75 µM), it reduces the reduction of the mass of fresh leaf produced by the metal and cancel its effect on the mass of dry matter without changing the reduction in the percentage of water content. No changes by PAs inhibitory effects of fresh and dry masses due to cadmium on the stems and roots were observed. Only the water content of these organs, reduced by cadmium, was restored to the value of the control by the PAs.

The extent of pigment used to evaluate the toxicity of metals. The Cd exposure for 84 h led to a significant decrease of chlorophyll a. This effect is consistent with earlier work (Padmaja et al. 1990; Larsson et al. 1998; Groppa et al. 2007a, b). and follows, according Mysliwa-Kurdziel and Strzalka (2002), an inhibition by Cd, the biosynthesis of chlorophyll or activation of chlorophyllase (Abdel-Basset et al. 1995). Cadmium does not alter the content of chlorophyll b and carotenoids and, again, chlorophyll b apprears to be less sensitive than chlorophyll a to Cd stress. At the toxicity of Cd, a slight lowering of the ratio of chl a / b, agreement with the work of Baszynski et al. (1980) and Larsson et al. (1998). The application of exogenous Put or Spd does not alter the levels of pigments. However, when the foliar spray of one or other of PAs was associated with the treatment by cadmium, it helps to keep chlorophyll a and, in the case of Spd, to increase the content of chlorophylls a and b. The PAs are present in the membranes of thylacoides, the PSII and light-harvesting complex II (Kotzabatsis et al. 1993; Legocka and Zajchert, 1999) and known for their antisennescent effect (Galston and Kaur-Sawhney, 1987; Pandey et al. 2000). The fact that the only exogenously application of Put or Spd didn't produce any effect on pigments suggests that the Put and Spd foremost protect the pigments by a trapping of ROS produced under the effect of Cd, ROS that lead to the destruction pigments.

Indeed, many works related a protective effect of exogenous PAs against oxidative stress generated by another environnemental factors such ozone (Bors et al. 1989; Navakoudis et al. 2003), UV radiation (Sfichi et al. 2004), paraquat (Chang and Kao, 1997; Benavides et al. 2000), acid rain (Velikova et al. 2000), high salinity (Tang and Newton, 2005), water deficit and chilling (Nayyar and Chander, 2004), and metals (Groppa et al., 2001, 2007 a, b). Furthermore, the protection of the photosynthetic apparatus by exogenous Put against ozone and UV radiation has been reported by Navakoudis et al. (2003) and Sfichi et al. (2004).

The results showed a Cd absorption by plant roots exposed for 84 h to $CdCl_2$ (75 µM) followed by transportation to parts, indicating transport by xylem as shown by salt et al. (1995) and Gong et al. (2003). The application of the Put did not alter the Cd content of various organs but a slight increase in the amount of Cd was observed in the leaves, because an increase by the diamine of the dry mass of these organs.

The spraying Spd changes the distribution pattern of Cd between different parts of the plant. Thus, Spd induced a very strong stimulation of root uptake followed by an increase of transport in the stems and this is true that the results are expressed in μg $Cd.g^{-1}$ D.W. or in μg Cd per organ. A slight decline in Cd was produced by the Spd in the leaves could explain the higher content of chlorophyll (less production of ROS and trapping by Spd). However, given the increase by the Spd of the leaf D.W., the amount of Cd accumulated was ultimately identical to that found in untreated plants to the Spd.

In an attempt to explain the effect of the root-level, Spd can assume a transport by the phloem of Spd to these bodies, transport demonstrated by earlier work (Béraud et al. 1992; Antognoni et al. 1998). A change in membrane permeability could arise and it is known that polyamine can influence the root absorption of a cation such a potassium (De Agazio et al. 1988). These results differ from those obtained by Hsu and Kao (2007) who report a reduction in the absorption of Cd by detached leaves of rice, after pretreatment with the Spd. This simplified system can probably explain this difference in effect observed.

It is now well known that abiotic stress including heavy metals cause molecular damage caused by overproduction of reactive species of oxygen (ROS). The overproduction of ROS cause lipid peroxidation which leads to the formation of degradation products such as alkanes and aldehydes (malondialdehyde) (Ferrat et al. 2003).

The malondialdehyde (MDA) which is a good marker for the lipoperoxidation, is significantly increased in the leaves of plants treated with cadmium. The foliar spary with either PAs can cancel (Put) or Sharply reduce (Spd) the production of MDA caused by the metaland the implementation of Put or Spd alone did not have any impact on the formation of this aldehyde. The results therefore reflect oxidative stress generated by cadmium and show a protective effect of PAs against metal stress. These data are consistent with the work of Groppa et al. (2001), Wang et al. (2007) and Zhao and Yang (2008) who report a reduction of lipid peroxidation induced by Cd and Cu by the exogenous application of Spd and Spm, respectively in leaf discs of *Helianthus annuus* and in the leaves of *Nymphoides peltatum* and *Malus hupehensis*.

To fight against oxidative damage, plants set up systems for defense and non-enzymatic antioxidant enzyme that plays a role in regulating the levels of ROS. Among the ensymes for the destruction of radical species include peroxidase.

A sharp increase in the activity of the guaiacol peroxidase (GPOX) is induced in leaves by treatment with Cd and this suggests a role for this enzyme in the elimination of $H_2O_2$ produced in excess. Several studies have suggested the involvement of guaiacol peroxidase in oxidative stress imposed by metals (Shah et al. 2001; Schützenbüdel and Polle, 2002; Sirvastava et al. 2004; Smeets et al., 2005; Mishra et al. 2006; Hsu and Kao, 2007; Semane et al. 2007). The foliar application of Spd or Put combined with Cd treatment can mitigate the increased acivity of guaiacol peroxidase induced by the metal. Contrary to what is observed by Tang et al. (2005), the only application of exogenous Put and Spd produces no change in the peroxidase activity, we can not conclude to an induction by Spd of *de novo* synthesis of this enzyme. It is more likely that exogenous Pas (Spd, Spm) have antioxidant function in protecting the tissues of oxidative damage induced by copper and cadmium as reported by Groppa et al. (2007 a, b) and Wang et al. (2007).

Exposure to heavy metals is often accomplished by a synthesis of various metabolites that accumulate in concentraions of the order of millimoles, which include some amino acids (Costa and Spitz, 1997; Sharma and dietz, 2006; Lei et al. 2007), especially proline, carbohydrates (Samarakoon and Rauser, 1979; Jha and Dubey, 2004), and polyamines (Taulavuori et al. 2005; Sharma and dietz, 2006; Groppa et al. 2007 a, b; Groppa and Benavides, 2008).The exposure of plants to cadmium produced a doubling of the content of total free amino acids in leaves. The foliar application of Put or Spd, whether conducted with or without treatment with Cd, did not affect the amino acid content of leaves. The accumulation of amino acids in response to metal stress can be explained by the degradation of certain proteins or by the *de novo* synthesis of amino acids (Berlet and Stadtman, 1997; Siedlecka and Krupa, 2002; Hsu and Kao, 2003; Pena et al. 2006; Sharma and Dietz, 2006).

Among the amino acids, proline, while constituting less than 5 % of total free amino acids (Matysik et al. 2002; Lei et al. 2007), is probably one of the most common metabolites that is synthetized in response to stress. Its content, which constitued 4.40 % of total amino acids, was significantly increased in leaves after treatment with cadmium but was unaffected by the only foliar application of Put or Spd. However, When the foliar application of one or other of PAs was associated with Cd treatment, it suppressed (with Put), or reduced (with Spd) the proline accumulation induced by Cd.

An increase in proline under the effect of cadmium has been reported for various plants such as lupine (Costa and Spitz, 1997), rice (Shah and Dubey, 1998), radish (Chen et al. 2003) and soybean (Balestasse et al. 2005). Such an increase in proline was also observed under the effect of other metals such as Mn (Lei et al. 2007), Zn, Pb, Co, Cu (Alia et al. 1995; Schat et al. 1997; Sharmila and Pardha Saradi, 2002; Sharma and Dietz, 2006), but also with other types of stress: saline (Demiral and Turkana, 2005), water (Taulavuori et al. 2005), UV radiation (Pardha Saradi et al. 1995), thermal (Taulavuori et al. 2005). Proline is considered part of a general adaptation syndrome to unfavorable environmental conditions. Different functions are attributed to the accumulation of fluid compatible osmoregulation, chelation and detoxification of metals, protection of enzymes, regulating cytosolic acidity, stabilizing the machinery of protein synthesis and trapping of reactive oxygen species (hydroxyl radicals, singlet oxygen) (Sharmila and Pardha Saradi, 2002).

The results of treatment with Cd combined with foliar aplication of Spd suggest that the proline could be as osmoticum or trapping of ROS. Indeed, a slight decrease in the percentage of water / D.W. accompanied by an increase in proline is observed, suggesting a role of osmoregulation of proline. However, the combination of two treatments (Cd and Put) did not affect the percentage of water / leaf D.W. or their content of proline (identical to that of control). In addition, the foliar spray of Spd partially reduced the production of MDA and proline induced by exposure to cadmium. This effect is consistent with a protective membrane against the attack of free radicals, agreeing with the results of Smirnoff and Cumbe (1989) which suggest that proline reacts with hydroxyl radicals to generate non-toxic hydroxyproline and those of Alia et al. (2001) and Matysik et al. (2002) who showed that proline was able to scavenge singlet oxygen. In addition, the increased content of chlorophylls could result from protection by proline of thylacoidal membranes against the attack of ROS as reported by Kavi Kishor et al. (2005).

In addition to changes in amino acid contents, accumulation of carbohydrates has been reported in response to different environmental stresses especially metals (Samarakoon and Rauser, 1979; Moya et al. 1993; Costa and Spitz, 1997; Dubey and Singh, 1999).

The results showed that the treatment with Cd produced a doubling of the soluble carbohydrate contents of leaves but the only foliar spray with Put or Spd did not change in these metabolites. However, the combination of two teatments: Cd-PA, permits with the Spd, a total abolition of the accumulation caused by Cd and only a mitigation with Put. This accumulation of carbohydrates as a result of Cd could, like what has been reported for arsenic by Jha and Dubey (2004), reflect a deterioration in enzyme activities including invertase acid, the sucrose synthase and starch phosphorylase. Costa and Spitz (1997) also observed a disruption of soluble carbohydrate under the effect of Cd resulting from the action of free radicals produced. An increase in photosynthesis appears to be unlikely to reflect the increase in soluble sugars observed, Cd producing a reduction of pigment. The foliar application of Put and Spd to plants exposed to Cd, resulting in the maintenance of pigments and therefore likely to photosynthesis. It can therefore be assumed by a railroad polyamine, free radicals produced under the effect of Cd, preventing (With Spd) or mitigating (With Put) and the disruption of carbohydrate metabolism.

Many studies showed the involvement of polyamines in defence mechanisms during the botic and abiotic stresses (Flores, 1991; Galston et al. 1997; Bouchereau et al. 1999; Taulavuori et al. 2005; Groppa and Benavides, 2008), and can operate according to Rhee et al. (2007) as stress molecules.

The exposure of plants to Cd product, in leaves, an increase of PA contents, free and conjugated forms together, the effect is especially marked for Put. These results agreed with previous results obtained under the effect of Cd and Cu for leaf tissues of wheat, oats, and bean that show an increase in free Put due to the stimulation of ADC (Weinstein et al. 1986; Groppa et al. 2003, 2007 a, b) and also that of ODC (Groppa et al. 2003, 2007 a, b). Moreover, a reduction of diamine oxidase activity by Cd and Cu, as reported by Groppa et al. (2003) and Balestrasse et al. (2005), could help to increase the content of free Put. An elevated content of conjugated Put was also reported by Weinstein et al. (1986) suggesting that conjugated PAs can be used like forms of storage and can be exchanged with the free forms. The treatment of plants by Cd product, however, reduced levels of free and conjugated PAs in the roots. The lowering of free PA forms was linked to the reduction of Put and more to the reduction of Spd and Spm contents. The sharp rise compared Put / Spd that results from 1.08 to 2.00, evidenced by an inhibition of Spd synthase by Cd and therefore the content of Spd, causing therefore a reduction of the Spd synthesis. However, the lowering of root PAs by Cd, could result, not only of reducing its synthesis but also an inhibition of the transport by the phloem. Indeed, the accumulation of Pas observed i the leaves could be partly due to the inhibition of transport.

The decrease in root PAs combined results of a reduction of conjugated forms of the Spd and the Spm and agrees with the notion that conjugated PAs can constitute forms of storage. It can not reflect this reduction of conjugated PAs by Cd inhibition of their transport by phloem, contrary to free PAs, the conjugated PAs are transported either by the phloem or xylem (Antognoni et al. 1998).

The folair spray of Put, surprisingly, change a little te content of free Put of the different organs, with the exception of a slight increase in the leaves and roots. However, a slight increase of the content of conjugated Put was observed in the different parts of plant. These results suggested that conjugated Put could help to keep constant the contents of free Put as shown by Kuthanova et al. (2004). Treatment with the Put did not alters or a little the contents of Spd and Spm of different organs but reduced sharply the content of their conjugated forms in the leaves and roots. These data showed that exogenous Put did not alter the content of endogenous free Put, Spd and Spm , that may imply a lack of penetration of diamine due to its cationic properties which leads to bind wall anionic sites (Evans and Malmberg, 1989; Tiburcio et al. 1990). However, the foliar spray of Put causing in hand, an increase of conjugated Put, and in the other hand, a decrease of conjugated forms of Spd and Spm in the leaves and roots, indicated that, Put had not only been absorbed at foliar level but also transported through the phloem to the roots. Indeed, the absorption of PAs had been identified in various parts of plants (Theiss et al. 2004) without being able to conclude the existence of a high affinity general transport system as has been demonstrated for animal cells (Hubert et al. 1996; Seiler et al. 1996; Sakata et al. 2000). Furthermore, we know that PAs are transported by the xylem (Friedman et al. 585 1986; Antognoni et al. 1998). We can suppose that free PAs homeostasis can be maintained and associated with the disruption of conjugated forms.

The combination of Put and Cd led to keep the contents of total PAs at leaf levels similar to those of control leaves with a change in the distribution pattern of PAs: increase of Put and decrease of Spd and Spm. It therefore appeared that treatment with Put did not alter the effect of Cd on the foliar Put content but slightly lowered those of Spd and Spm than those of control. Moreover, the combination of Put and Cd resulted in a reduction (conjugated Put) or deletion (conjugated Spd and Spm) effect of Cd on the contents of foliar conjugated forms. The stem contents of total free and conjugated PAs were respectively increased by 28% and 15 % by combination treatment of Put and Cd, mainly because of an accumulation of free and conjugated Put.

Therefore, it appeared that the application of Put permitted som protection to leaves, reducing the stimulatory effects of Cd on their levels of free and conjugated Spd and Spm, but did not change the disruption caused by the metal alone on the levels of free and conjugated Put of leaves and those of free Put of roots.

The foliar spray of Spd produced an increase in total free PAs in the various organs, resulting from an increase in Spd and especially Put. This effect was particularly marked in leaf, because there was an increase of Spd, after a local spraying of this triamine but a doubling of the content of Put. Such a rise in Put has been reported by De Agazio et al. (1995) and Duhazé et al. (2002) after contribution of exogenous Spd whose degradation by polyamine oxidase generated Put; this allowed to adjust the excess of Spd. We can not exclude also the participation of an interconversion pathway of Spd in Put, caalyzed by spermidine / spermine N1-Acetyltransferase (SSAT), highlight among animal cells (Pegg, 1986; Rhee et al. 2007). Indeed, the contribution of exogenous acetylspermidine to germination of Arabidopsis thaliana, resulting in a sharp increase in free Put (Tassoni et al. 2000) and moreover, the presence of Spd and Spm was acetylated reported by Torrigiani et al. (1993) and Del Duca et al. (1995).

Treatment with Spd induced a decreased in total conjugated PAs in different parts of the plant, due to lower conjugated Put in aerial parts, and for stems and roots, to decrease of conjugated forms of Spd and Spm which can only exlplained by a reduce of the transferase activities to the combination with hydroxycinnamic acids (Sun et al. 619 1991).

When Spd was combined with Cd, a sharp increase of free PAs was observed in aerial parts of plants, resulting mainly from a very significant increase of free Put, significantly higher than those observed with the single application of Spd. This accumulation of Put could result both of stimulating the synthesis induced by the metal but also to the inteconversion of Spd in Put. At root levels, the combination treatment Spd-Cd greatly increased the content of Put opposing the inhibitory effect produced by Cd alone and amplified the action of treatment with exogenous Spd.

The combination treatment of Spd with Cd, will maintain conjugated PAs sheets at values similar to those found in control plants, negating the effect of Cd.

These results suggested that Spd might play a protective role with respect to the toxicity of metal. It was likely that the accumulation of Put observed in different parts of the plant resulted from an interconversion of Spd to Put, which can confer a certain tolerance to the stress induced by Cd.

**Acknowledgement**

We thank Dr Thierry Guinet from 'Ecole Nationale d'Enseignement Supérieure Agronomique (ENESA) (Dijon-France) for providing us the seeds of the spring line *Brassica juncea* AB79/1.

**Figure**

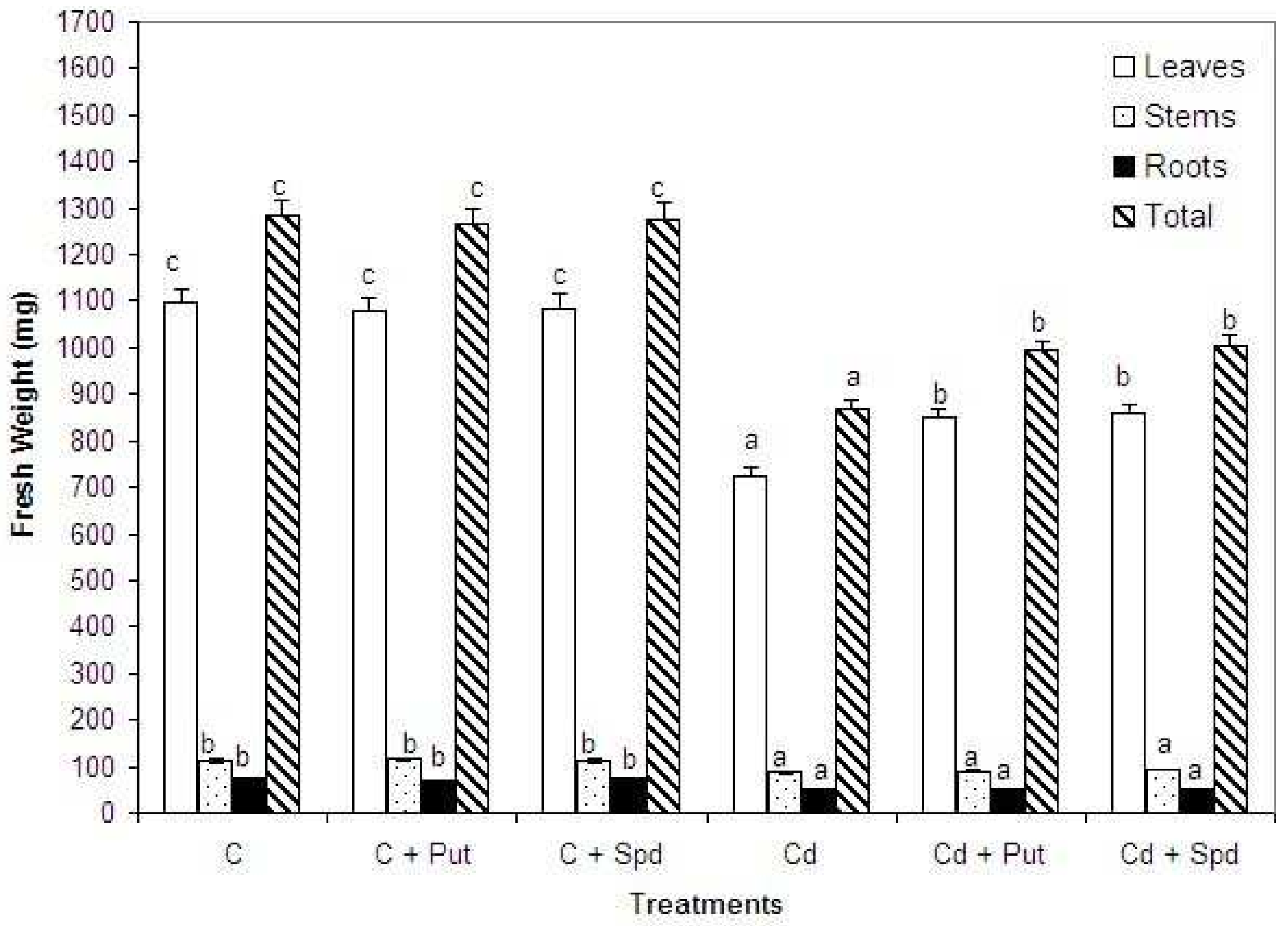

Figure 1: Effect of an exogeneous application, by spraying, of either Put (1 mM), or Spd (1 mM), on the fresh weight of organs from *Brassica juncea* plants grown under hydroponic conditions and exposed to $CdCl_2$ (75 µM) for 84 h. The results, expressed as means ± S.D., were calculated from 3 independent experiments made in triplicate and conducted on, at least, 30 plants per experiment. For each organ, the different letters indicate differences significant at $p = 0.05$ (One-way ANOVA and LSD test).

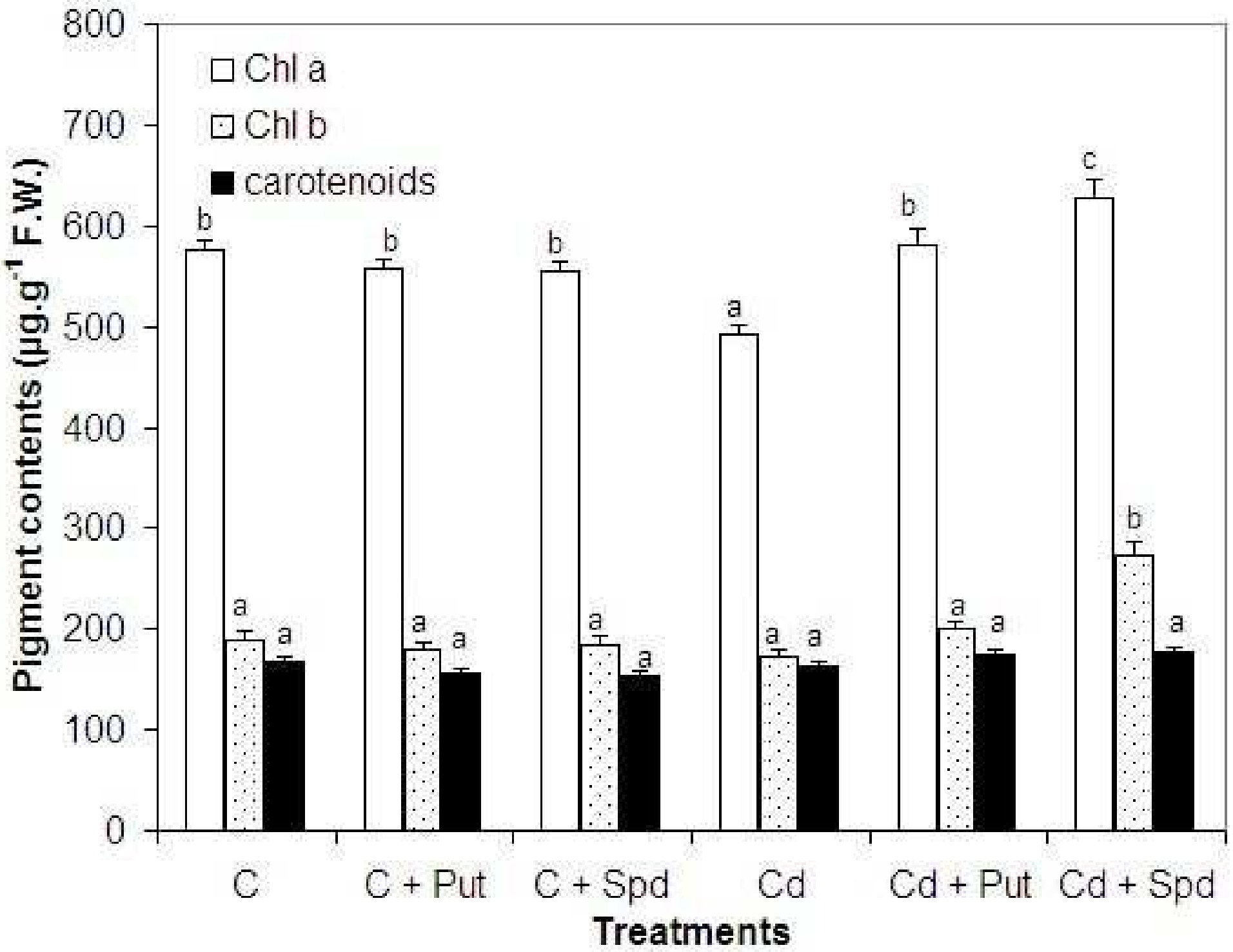

Figure 2: Effect of an exogeneous application, by spraying, of either Put (1 mM), or Spd (1 mM), on the Chl a, Chl b and carotenoid contents of leaves from *Brassica juncea* exposed to $CdCl_2$ (75 µM) for 84 h. The results, expressed as means ± S.D., were calculated from 3 independent experiments made in triplicate. The different letters indicate differences significant at $p = 0.05$ (One-way ANOVA and LSD test).

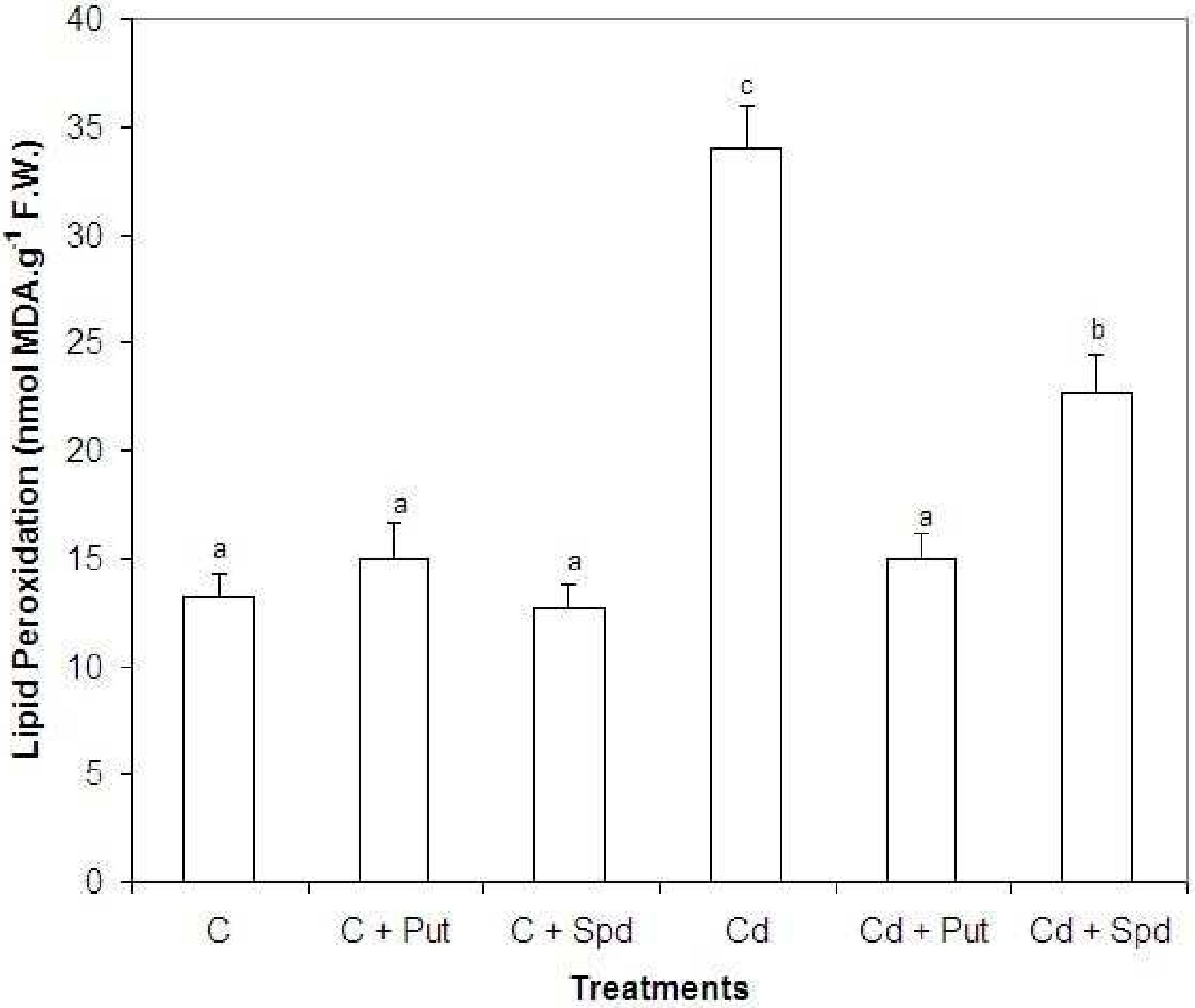

Figure 3: Effect of an exogeneous application, by spraying, of either Put (1 mM), or Spd (1 mM), on the MDA contents of leaves from *Brassica juncea* plants grown under hydroponic conditions and exposed to $CdCl_2$ (75 µM) for 84 h. The results, expressed as means ± S.D, were calculated from 3 independent experiments made in triplicate. The different letters indicate differences significant at $p = 0.05$ (One-way ANOVA and LSD test).

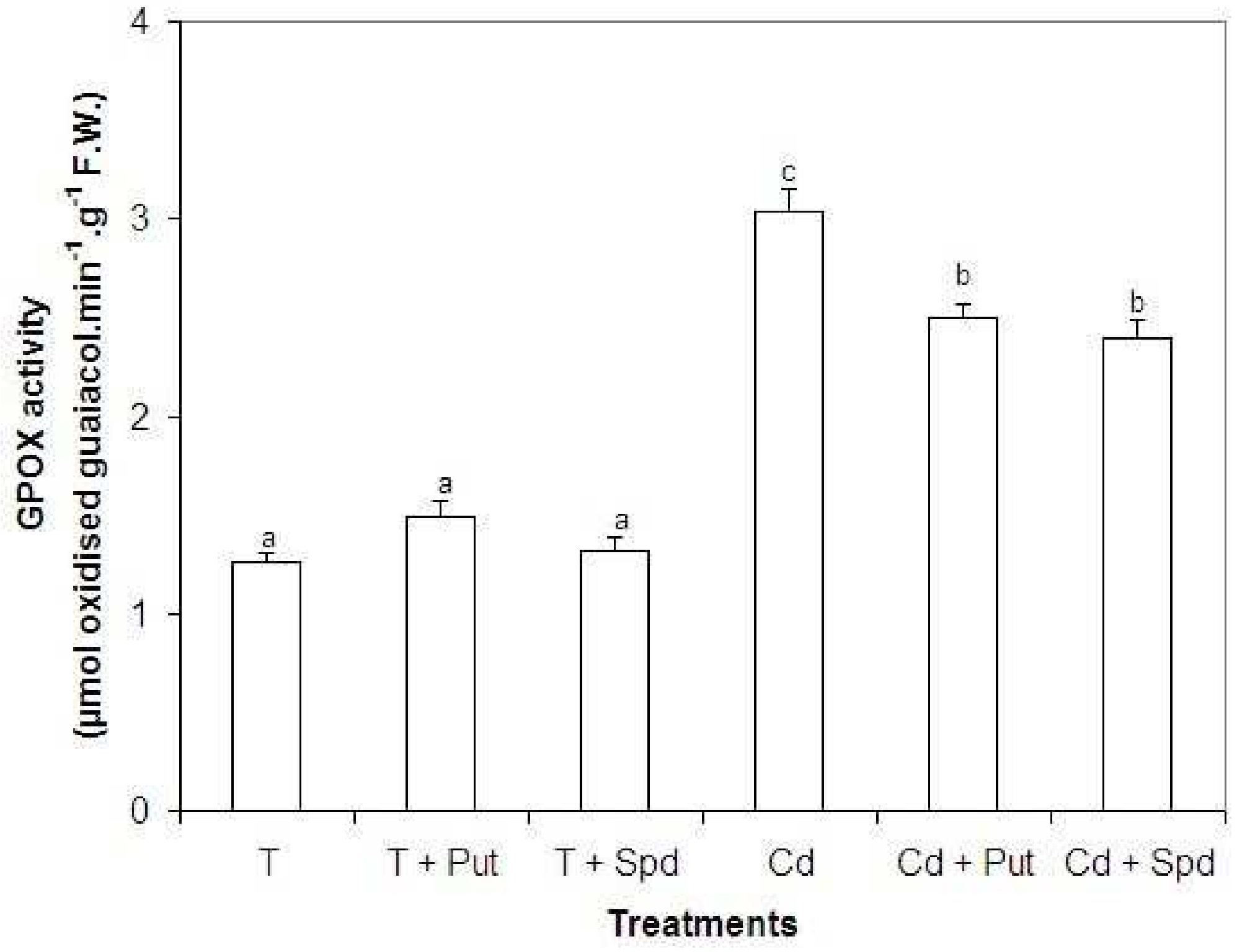

Figure 4: Effect of an exogeneous application, by spraying, of either Put (1 mM), or Spd (1 mM), on the guaiacol activity in the leaves of *Brassica juncea* plants exposed to $CdCl_2$ (75 µM) for 84 h. The results, expressed as means ± S.D, were calculated from 3 independent experiments made in triplicate. The different letters indicate differences significant at $p = 0.05$ (One-way ANOVA and LSD test).

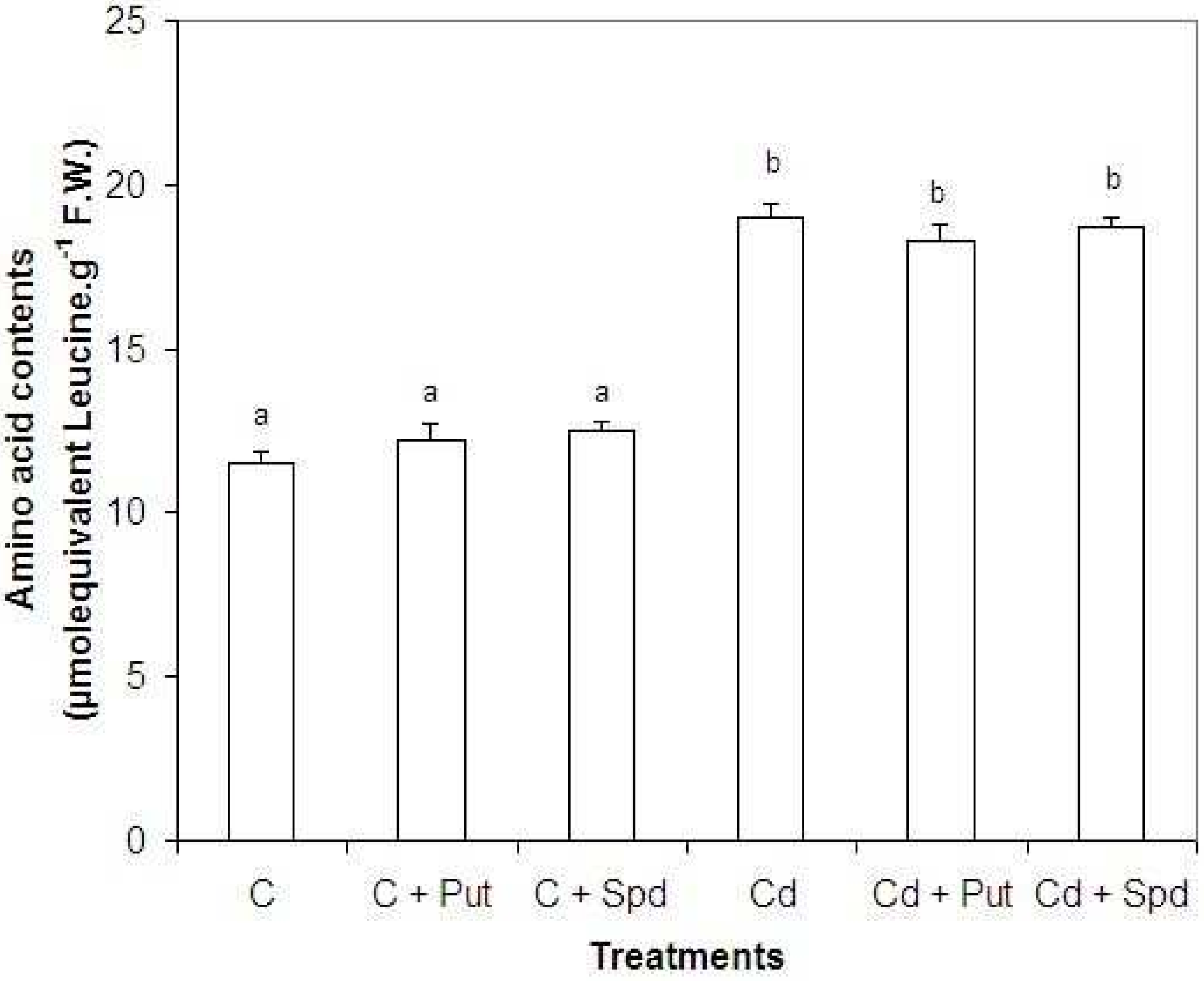

Figure 5: Effect of an exogeneous application, by spraying, of either Put (1 mM), or Spd (1 mM), on the amino acid contents of leaves from *Brassica juncea* plants grown under hydroponic conditions and exposed to $CdCl_2$ (75 µM) for 84 h. The results, expressed as means ± S.D, were calculated from 3 independent experiments made in triplicate. The different letters indicate differences significant at $p = 0.05$ (One-way ANOVA and LSD test).

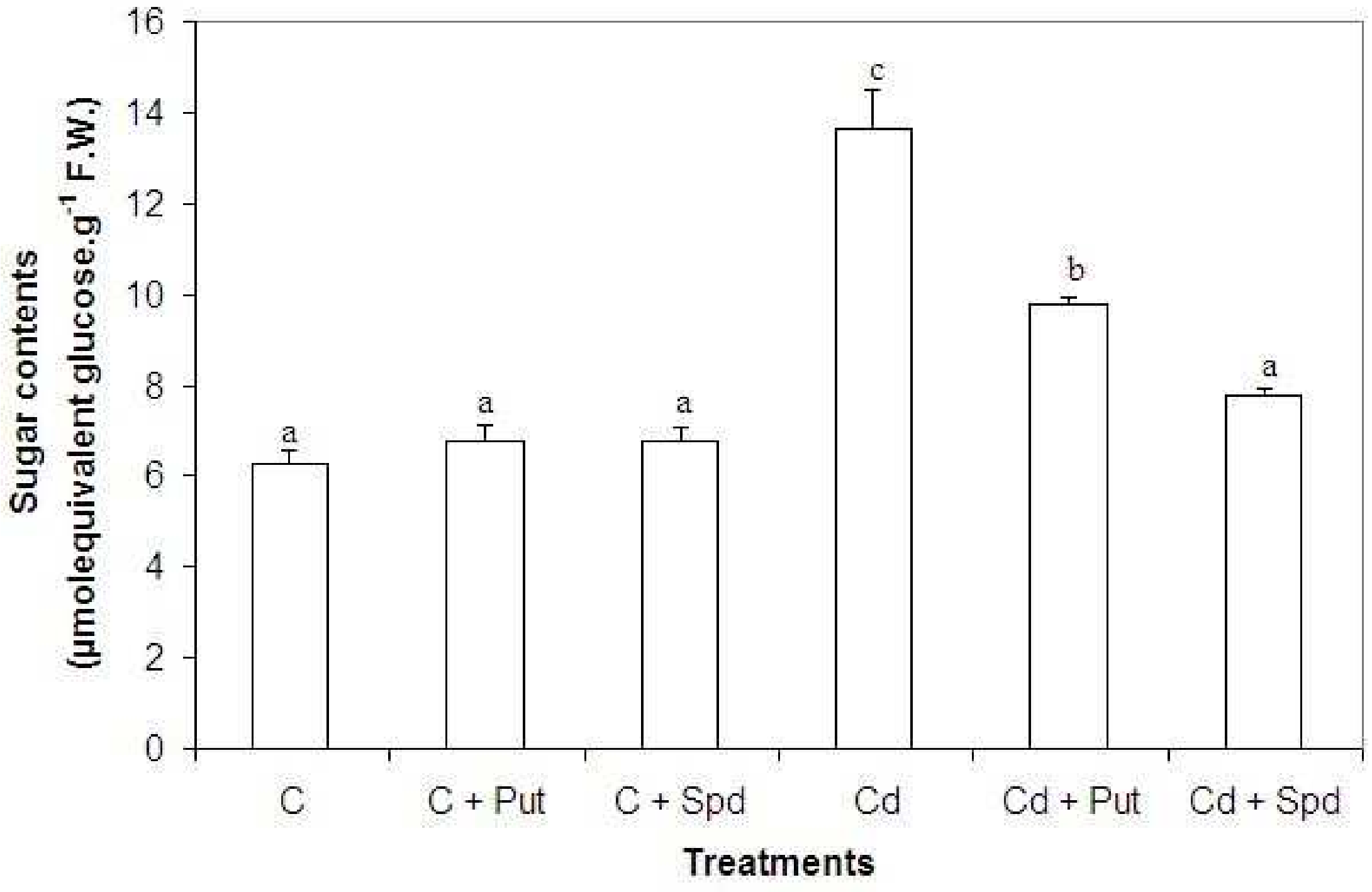

Figure 6: Effect of an exogeneous application, by spraying, of either Put (1 mM), or Spd (1 mM), on the soluble sugar content of leaves from *Brassica juncea* plants grown under hydroponic conditions and exposed to $CdCl_2$ (75 µM) for 84 h. The results, expressed as means ± S.D, were calculated from 3 independent experiments made in triplicate. The different letters indicate differences significant at $p = 0.05$ (One-way ANOVA and LSD test).

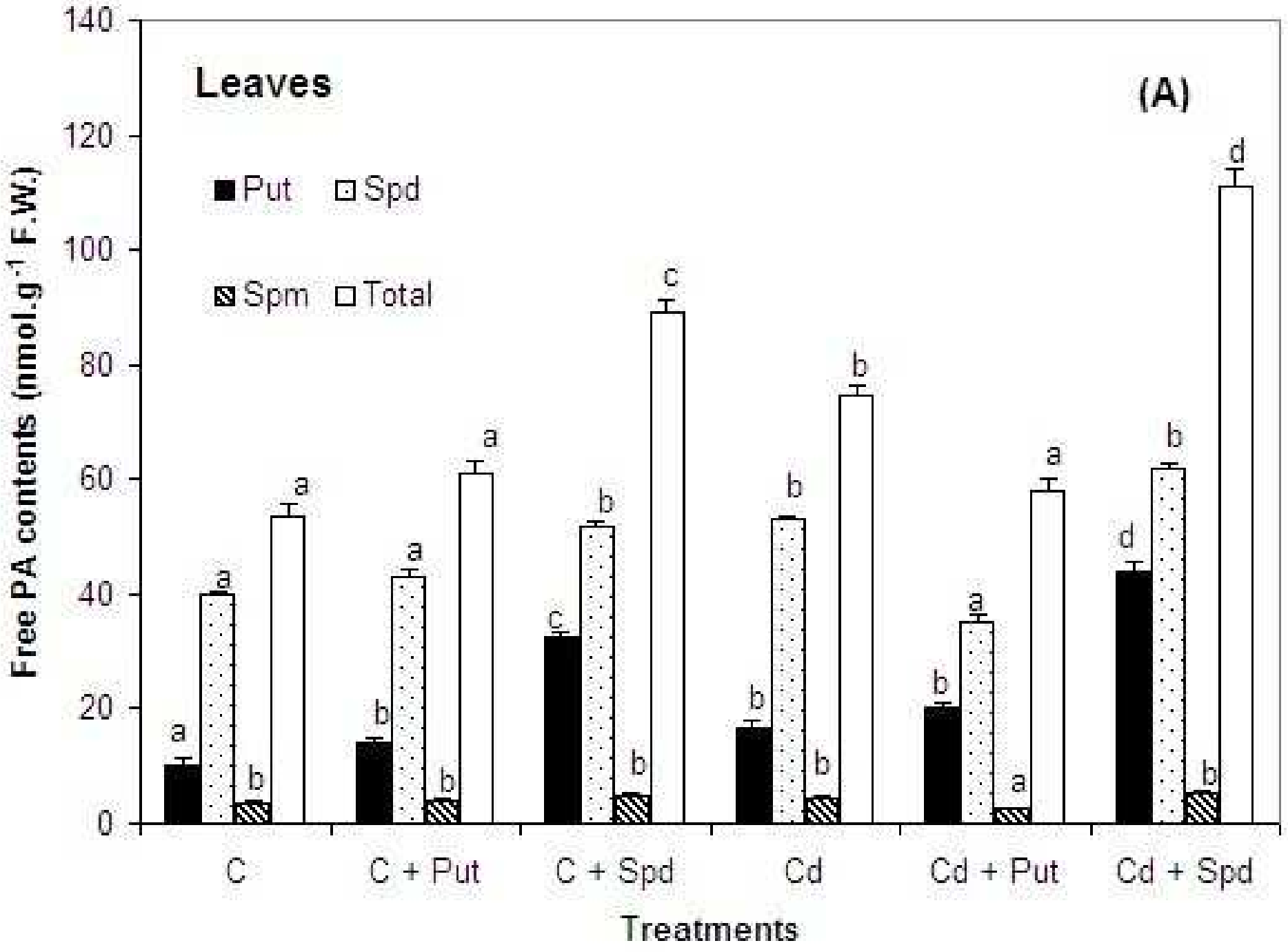

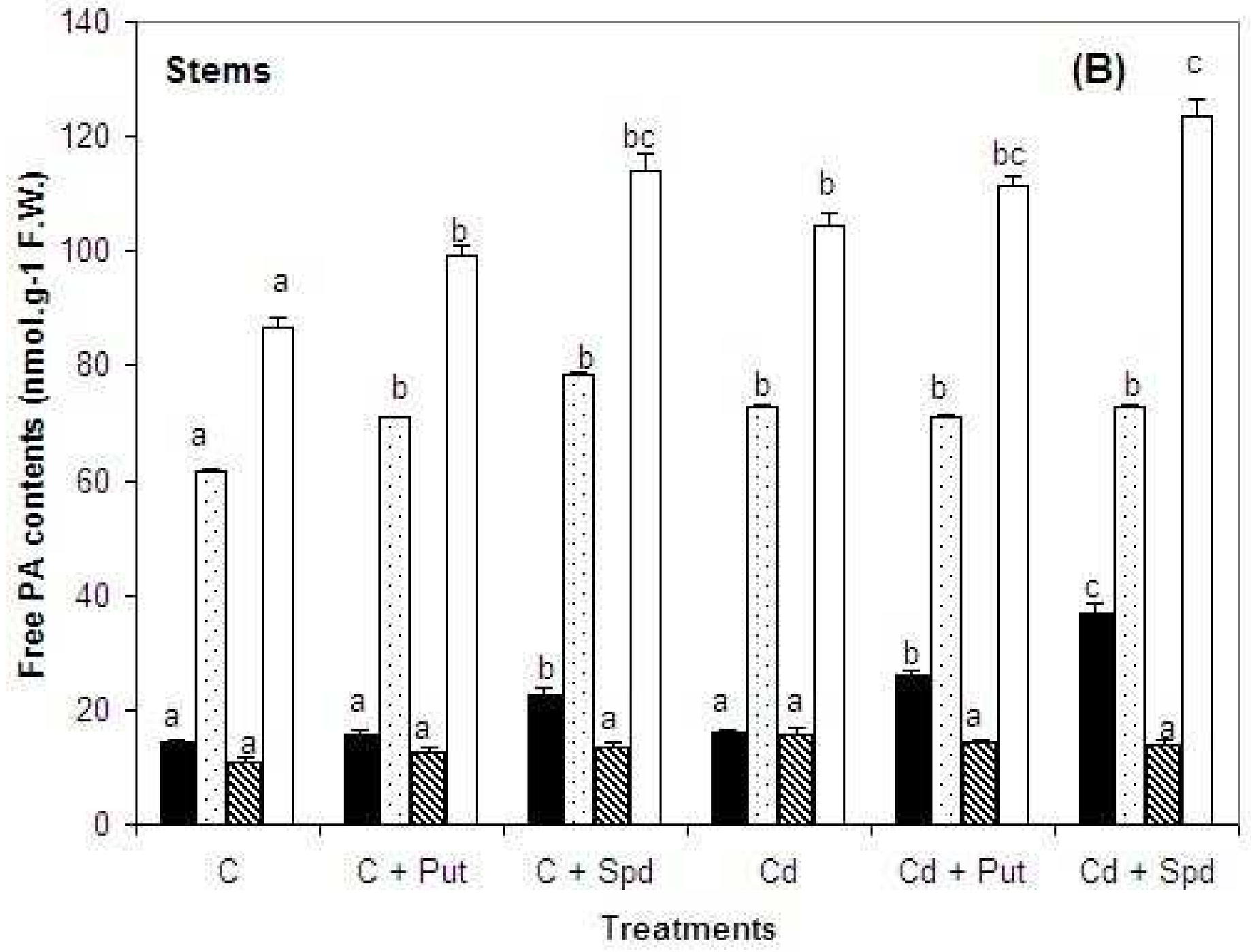

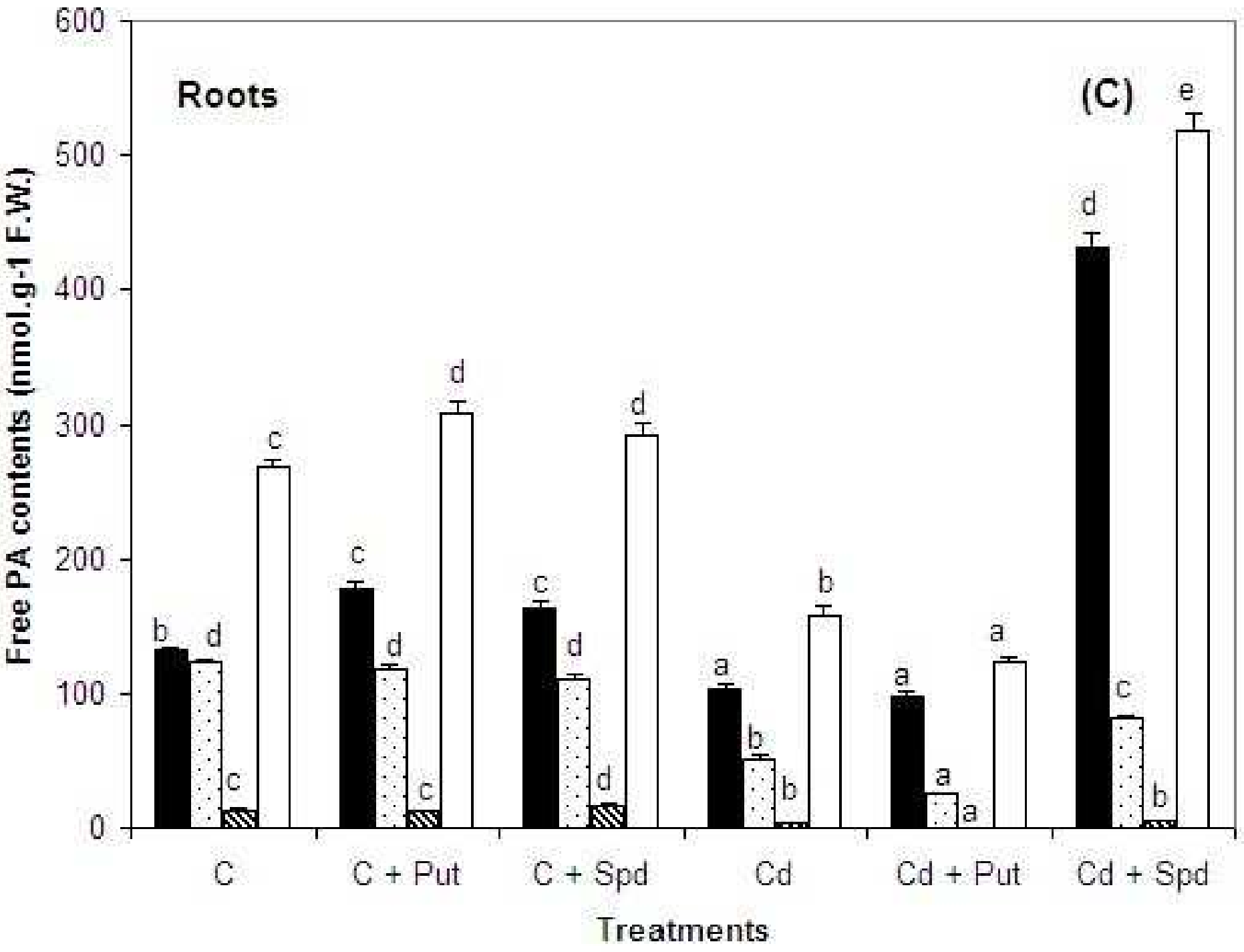

Figure 7: Effect of an exogeneous application, by spraying, of either Put (1 mM), or Spd (1 mM), on the free polyamine content of leaves: (A), stems: (B) and roots: (C) from *Brassica juncea* plants exposed to $CdCl_2$ (75 µM) for 84 h. The results, expressed as means ± S.D, were calculated from 3 independent experiments made in triplicate. The different letters indicate differences significant at $p = 0.05$ (One-way ANOVA and LSD test).

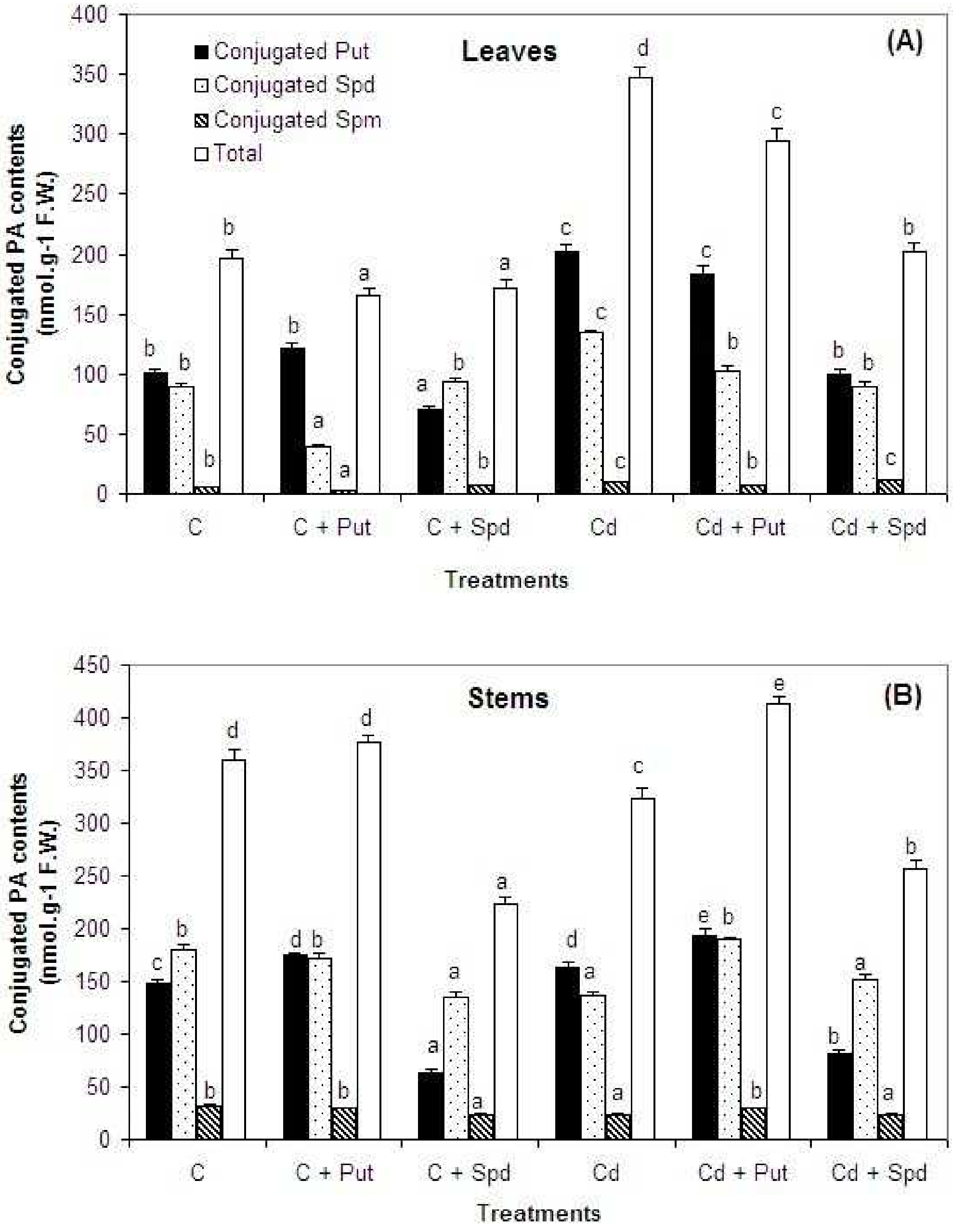

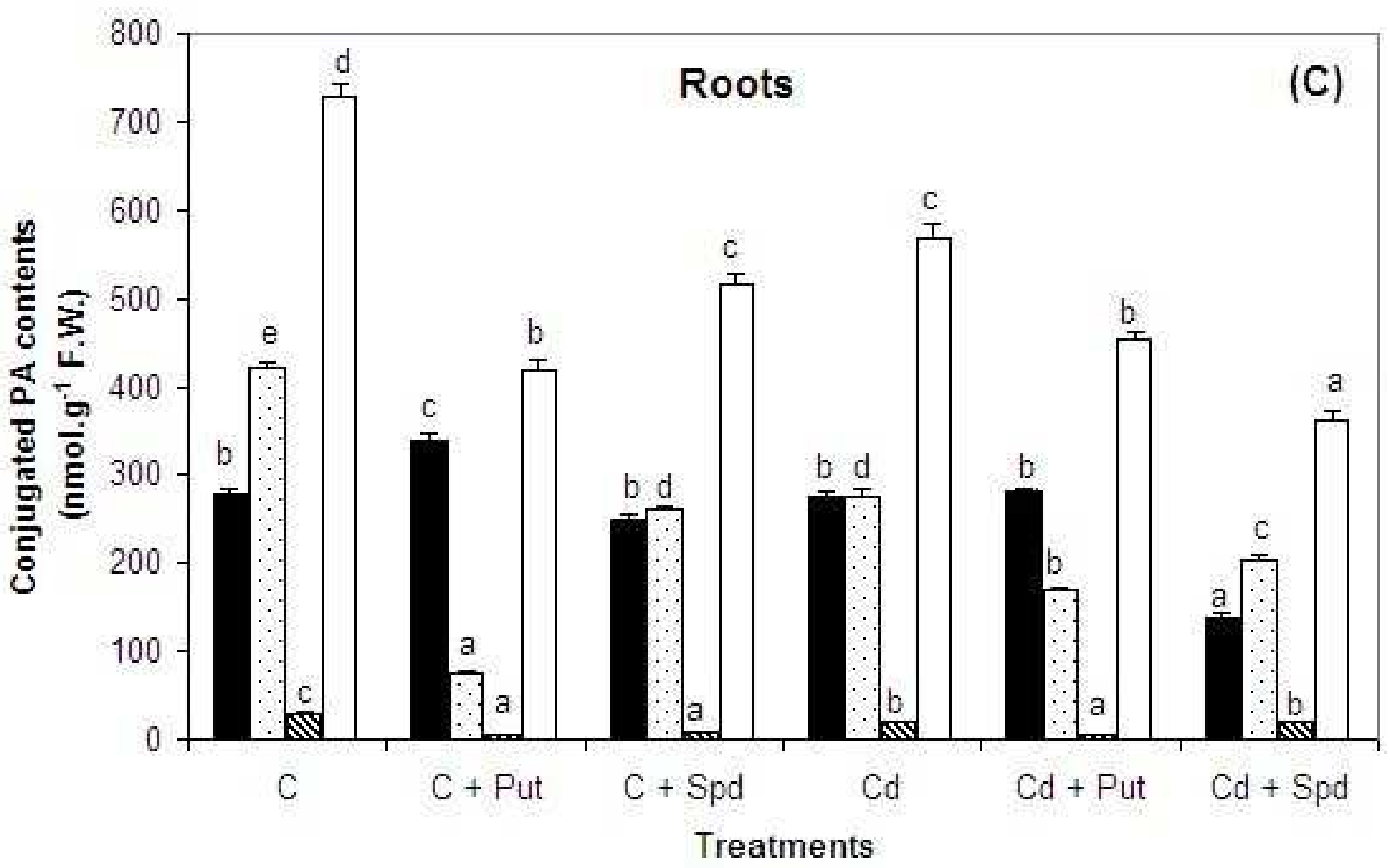

Figure 8: Effect of an exogeneous application, by spraying, of either Put (1 mM), or Spd (1 mM), on the conjugated polyamine content of of leaves: (A), stems: (B) and roots: (C) from *Brassica juncea* plants exposed to $CdCl_2$ (75 µM) for 84 hours. The results, expressed as means ± S.D, were calculated from 3 independent experiments made in triplicate. The different letters indicate differences significant at $p = 0.05$ (One-way ANOVA and LSD test)

Table
Click here to download Table: Tables.doc

Table 1: Effect of an exogeneous application, by spraying, of either Put (1 mM), or Spd (1 mM), on the dry weight of organs of *Brassica juncea* seedlings grown under hydroponic conditions and exposed to $CdCl_2$ (75 µM) for 84 h.

| Treatments | Organs | | |
|---|---|---|---|
| | Leaves | Stems | Roots |
| C | $96.13 \pm 3.03^{b}$ | $12.73 \pm 0.40^{e}$ | $11.69 \pm 0.62^{ef}$ |
| C + Put | $110.94 \pm 2.26^{a}$ | $14.56 \pm 0.40^{d}$ | $11.74 \pm 0.50^{ef}$ |
| C+ Spd | $118.92 \pm 2.58^{a}$ | $14.73 \pm 0.51^{d}$ | $14.12 \pm 0.51^{d}$ |
| Cd | $79.47 \pm 1.99^{c}$ | $11.66 \pm 0.60^{ef}$ | $9.69 \pm 0.44^{g}$ |
| Cd + Put | $95.95 \pm 2.29^{b}$ | $10.93 \pm 0.34^{f}$ | $8.60 \pm 0.37^{g}$ |
| Cd + Spd | $97.67 \pm 2.54^{b}$ | $11.05 \pm 0.26^{f}$ | $8.59 \pm 0.18^{g}$ |

The results, expressed as means ± S.D, were calculated from 3 independent experiments made in triplicate. The different letters indicate significant differences at $p = 0.05$ (One-way ANOVA and LSD test).

Table 2: Effect of an exogeneous application, by spraying, of either Put (1 mM), or Spd (1 mM), on Cd accumulation by organs of *Brassica juncea* plants grown under hydroponic conditions and exposed to $CdCl_2$ (75 µM) for 84 h.

| Treatments | Organs | | |
|---|---|---|---|
| | Leaves | Stems | Roots |
| C | 24.19 ± 1.26$^{g}$ | 22.11 ± 2.07$^{g}$ | 21.67 ± 2.47$^{g}$ |
| C + Put | 23.24 ± 1.23$^{g}$ | 22.78 ± 2.30$^{g}$ | 22.50 ± 2.18$^{g}$ |
| C + Spd | 23.33 ± 1.18$^{g}$ | 22.89 ± 2.18$^{g}$ | 22.83 ± 3.47$^{g}$ |
| Cd | 355.24 ± 22.31$^{e}$ | 932.67 ± 63.57$^{d}$ | 4506.00 ± 307.98$^{b}$ |
| Cd + Put | 336.67 ± 26.67$^{e}$ | 943.33 ± 64.67$^{d}$ | 4771.83 ± 336.48$^{b}$ |
| Cd + Spd | 285.09 ± 15.13$^{f}$ | 1139.67 ± 57.94$^{c}$ | 8378.33 ± 459.39$^{a}$ |

The results, expressed as means ± S.D, were calculated from 3 independent experiments made in triplicate. The different letters indicate differences significant at $p = 0.05$ (One-way ANOVA and LSD test).

Table 3: Effect of an exogeneous application, by spraying, of either Put (1 mM), or Spd (1 mM), on the proline content of leaves from *Brassica juncea* plants grown under hydroponic conditions and exposed to $CdCl_2$ (75 µM) for 84 h.

| Treatments | Proline Contents (µmol.g$^{-1}$ M.F.) | % |
|---|---|---|
| C | 0. 55 ± 0.02$^{a}$ | 100 |
| C + Put | 0.592± 0.04$^{a}$ | 107 |
| C + Spd | 0.56 ± 0.04$^{a}$ | 101 |
| Cd | 1.26 ± 0.06$^{c}$ | 228 |
| Cd + Put | 0.63 ± 0.03$^{a}$ | 113 |
| Cd + Spd | 0.91 ± 0.05$^{b}$ | 165 |

The results, expressed as means ± S.D, were calculated from 3 independent experiments made in triplicate. The different letters indicate differences significant at $p = 0.05$ (One-way ANOVA and LSD test).